# Experimental Analysis and Evaluation of RaptorQ Codes for Video Multicasting over Wi-Fi

Berna Bulut

***Abstract*—**This paper presents a reliable and efficient high quality video streaming solution for use in challenging outdoor environments over Wi-Fi. An application layer forward error correction based on RaptorQ codes was implemented in a practical Wi-Fi based server and client system to enhance reliability. Thus, this paper presents the first detailed analysis on the implementation of RaptorQ codes for streaming high definition video over Wi-Fi. The measurements were performed in central Bristol with parameters such as RaptorQ symbol size, code rate, buffering time and modulation and coding scheme, and user quality of experience based on these parameters was evaluated. For multicast live video streaming it is demonstrated that system performance is mostly dominated by hardware and software limitations on constrained host platforms where the incoming packet rate exceeds the device`s ability to consume the traffic, i.e., Wi-Fi clients are a major source of packet loss, even in ideal channel conditions. Client limitations were found to be a function of modulation and coding schemes and RaptorQ coding parameters. Therefore, the optimum system design parameters such as RaptorQ symbol size, code rate and buffering time with respect to modulation and coding schemes were suggested considering practical limitations from real-world measurements.

***Index Terms*—** Application layer forward error correction, multicast, RaptorQ codes, video streaming, Wi-Fi.

## I. Introduction

The widespread use of mobile devices such as smart phones and tablets has led to a growing demand for mobile multimedia applications. There are applications, e.g., news broadcast and video streaming in concerts, in which multiple users attempt to access the same content. When unicast transmission is used, a significant amount of bandwidth is required since the multimedia data are transmitted to each user separately. Therefore, the networks run out of bandwidth for the high user densities. The problem becomes severe when each unicast user also requests the retransmission of any lost packets using feedback channel. Although, this provides a reliable transmission, it prevents the dissemination of media rich content. One solution is to efficiently transmit bandwidth hungry applications such as video over error-prone wireless channels to multiple interested users in the form of a multicast service. However, the legacy IEEE 802.11 standard [1] does not provide any standardised solution for robust and reliable multicast transmissions. Multicast packets are delivered as a simple broadcast service in which Medium Access Control (MAC) layer packet retransmission cannot be implemented. Hence, multicast transmission results in high packet losses. In practice, to improve reliability, the more robust Modulation and Coding Scheme (MCS) is selected for multicast transmission at the expense of throughput to address poor channel conditions.

Multimedia data, especially video, are very sensitive to errors, latency and jitter which cause serious degradation in the received video quality. Therefore, in order to ensure a high Quality of Experience (QoE), it is necessary to employ advanced and innovative error control mechanisms, which offered robust video multicasting. To this end, Application Layer Forward Error Correction (AL-FEC) based on Raptor codes [2] is standardised by 3rd Generation Partnership Project (3GPP) over Multimedia Broadcast and Multicast Services (MBMS) [3] to provide robust and reliable multicast/broadcast services over unreliable channels. AL-FEC enables more efficient MCS modes to be selected for multicast transmission. However, AL-FEC reliability comes at the expense of additional bandwidth requirements, hence the amount of redundant data (or code rate) should be determined based on the wireless channel conditions and the selected MCS in order to utilise the valuable radio and network resources most efficiently, i.e., there is a need for a cross-layer approach where the MCS and Raptor code parameters are jointly selected depending on the channel conditions. The 3GPPP selected Raptor codes due to their higher performance compared to the existing AL-FEC schemes, i.e., Raptor codes are implemented in software rather than hardware. Furthermore, the amount of redundant data can be determined on-the-fly thus precise knowledge of the channel does not required before the data transmission. There are available two practical Raptor codes: the 3GPPP standardised Raptor 10 (R10) [2] and the latest version of Raptor codes called RaptorQ (RQ) which provides exceptional protection performance and encoding parameters [4].

Therefore, this paper presents experimental results and analysis of AL-FEC based on the latest RQ codes for transmitting high data rate live video over Wi-Fi in real world outdoor environments. The main contributions of this paper are to use a customized Access Point (AP) to select higher MCS modes for high quality video multicasting, which is not common in the legacy IEEE 802.11 as the lowest and the robust MCS mode is used for multicast transmission, evaluate and hence provide comprehensive information and guidance on the cross-layer design and implementations of the RQ codes, investigate end user performance in a real-world environment and finally present the practical problems and system limitations on reliable and scalable live video multicasting over Wi-Fi. Here, a full set of design parameters such as RQ code rate and symbol size based on practical tablet based measurement campaigns conducted on the streets of Bristol, UK are recommended.

The rest of the paper is organized as follows. Section II provides the most important related works. Section III provides an overview of Raptor code AL-FEC. Section IV explains the measurement setup and parameters. Conducted experimental results and analysis are presented in Section V. Finally, Section VI concludes the paper.

## II. Related Work

The AL-FEC protection based on Raptor codes has been extensively studied for broadcast and multicast services over wireless networks. Specifically, these works investigated the system trade-offs between AL-FEC and physical layer parameters (e.g., MSC) for streaming services over WLANs [5], [6], Long Term Evolution (LTE) [7], [8] and Digital Video Broadcasting – Handheld (DVB-H) [9] and showed that only a well-designed and optimised system can maximise the spectral efficiency and user QoE. All these works evaluated the system performance and provided the optimum system parameters using theory and simulations. However, as stated in [10] that applying the theoretical formulas and/or using simplified simulations fail to match the actual user performance (user QoE), i.e., assuming independent and random packet losses in the channel, where the instantaneous variations of the wireless channel are not considered, leads to inaccurate estimation of Raptor codes` performance since Raptor code rate (the amount of redundant data) is defined according to channel loss in order to reduce FEC data overheads.

Note that all these works considered the cross-layer optimisation based on the 3GPP standardised legacy R10 codes [2], however, there exists a latest and more practical version of the Raptor codes which is called RQ [4]. Although, R10 codes have been extensively studied in the literature, there are few works on RQ codes. In [11] the performance of RQ codes in MBMS download and streaming services was evaluated and compared with the legacy R10 codes. They concluded that the performance of RQ codes is very close to an ideal FEC code since the minimised required additional data enables RQ operating with significantly lower transmission overhead compared to the standardised R10 codes. A cross-layer design and optimisation of RQ codes for streaming services over WLANs was presented in [12], [13]. The work in [14] provided a performance evaluation of RQ codes for streaming services over LTE and compared its performance with the legacy R10 codes. All these works again used simulations to evaluate the system performance. Only work which implemented RQ codes in a practical Wi-Fi based server/client system and reported its performance in terms of real world measurements for download delivery was presented in [15]. To do best of the author knowledge there is no work which implemented RQ codes for streaming services. Therefore, this paper presents the first detailed analysis on the implementation of RQ codes from real world measurements for streaming services over Wi-Fi.

## III. OVERVIEW OF RAPTOR CODES

FEC codes are applied cross packets at the application layer to overcome packet losses and hence provide reliable multimedia services over wireless networks. Raptor codes are a class of fountain code that can generate, in theory, an infinite number of encoded symbols from a source block with finite number of source packets on-the-fly. Due to this property, Raptor codes are characterized as rateless codes. In a fountain code, different users can make use of different received symbols to reconstruct the same data. Therefore, it does not matter which particular symbols are received, as long as the number of received symbols is enough to successfully decode the source block.

Raptor codes are very attractive due to low-complexity and flexibility. For example, a Raptor code processing requirement increases linearly with source block size $K$. This property enables Raptor codes to be implemented in software, which is uncommon for alternatives such as Reed-Solomon codes [16]. Furthermore, a source block size $K$ and the number of encoded symbols $N$ generated from a source block can be as large as desired. However, the 3GPP standardised R10 code, which is a systematic version where the original symbols are transmitted along with the encoded symbols, can support the coding parameters such as $4 \leq K \leq 8192$ and $K \leq N \leq 65{,}546$. The latest RQ has even more flexible coding parameters ($K$, $N$): it can support up to 56,403 source symbols in a single source block and generate up to $2^{24}$ encoded symbols from a fixed source block [4].

The main drawback of Raptor codes is that the decoder requires slightly more symbols than the original $K$ source symbols to decode a source block with high probability, i.e., the number of received symbols $R = K + \varepsilon$, where $\varepsilon$ denotes the reception overhead of Raptor codes which depends on $K$ and the desired probability that the source block can be fully reconstructed from the received symbol set [2]. RQ codes operate very close to ideal codes which has zero reception ($\varepsilon = 0$) overhead and has improved coding efficiency for different range of source block sizes compared with R10, which requires $K \geq 1000$ [3]. In practice using small block sizes is better for devices with limited power and processing capability such as mobile phones [17]. In this work at some points Raptor codes and RQ codes used interchangeable.

## IV. Measurement Setup

This section describes the measurement setup for RQ enabled High Definition (HD) video multicasting over Wi-Fi. First the implementations of the server and client systems will be described in detail and then the measurement scenario and the parameters will be given.

### *A. Server Configuration*

In this work, the video stream provided by the BBC (British Broadcasting Corporation) required a secure connection between the experimental server (as shown in Fig. 1) located on the University of Bristol' premises and the BBC server located in a remote location. A Secure Shell tunnel was established between the server and the raw Transport Stream [18] packets encapsulated in User Datagram Protocol (UDP) datagrams were delivered through the tunnel. The server was responsible for generating the RQ FEC packets in real-time from the incoming video packets (UDP datagrams) and used to send the resulting data stream via UDP broadcast for wireless transmission.

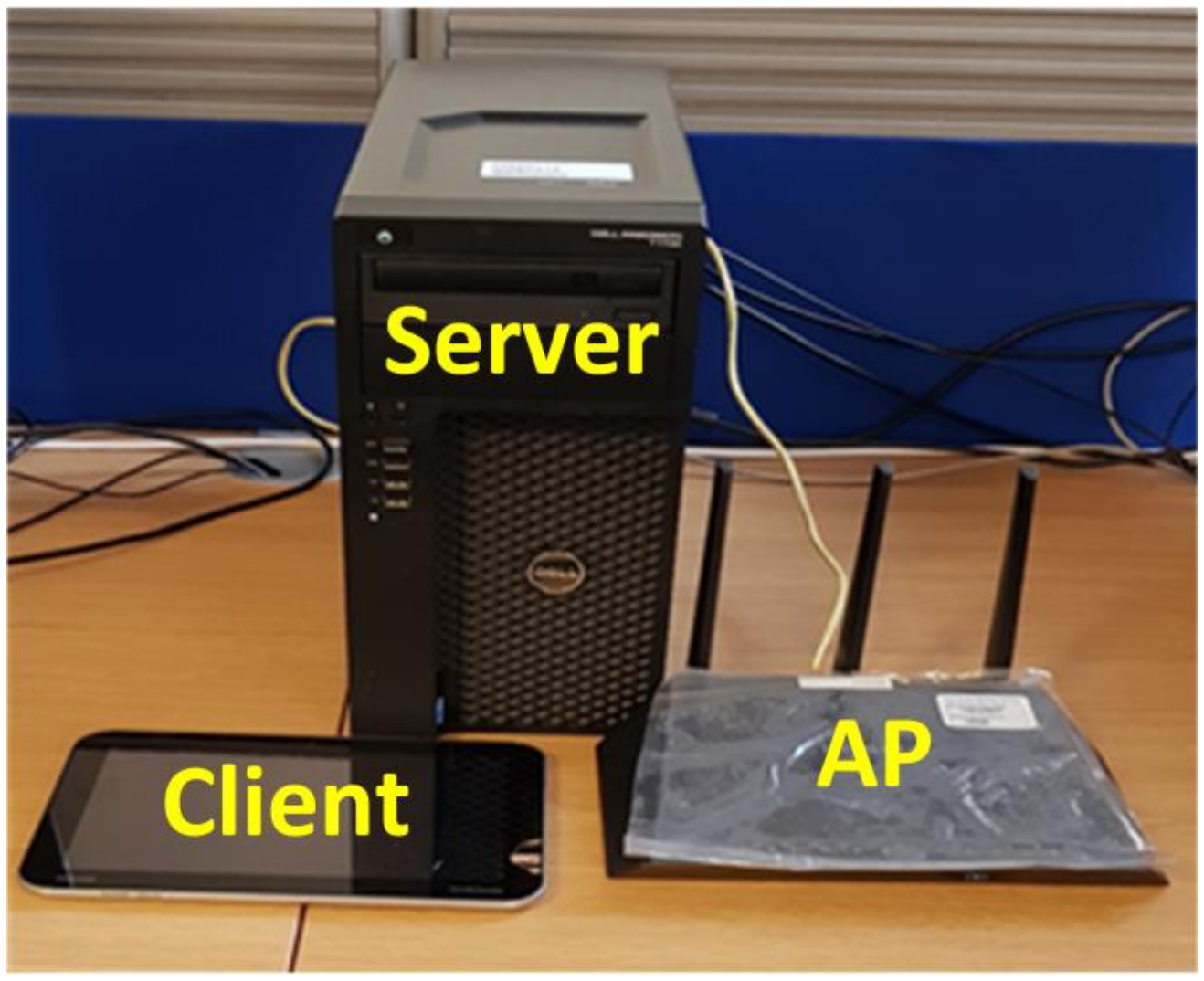

Fig. 1. Measurement setup: Server, AP and client.

The video streaming process is summarised in Fig. 2. All original video packets are prepended with their size and stored in a memory buffer as a stream. When the size of data in the buffer is larger than the desired RQ symbol size of $T$ bytes then: 1) a new coding symbol is formed from the first $T$ bytes of buffer, 2) $T$ bytes from the buffer are removed, 3) the size of tail from the previous packet is prepended to the next packet. When enough symbols are sent a new RQ block is formed and added to the memory storage for repair symbol generation. Any new video packets will be used to create the next RQ block. The RQ software, which runs in real-time on the server (encoder) and Android Tablets (decoders), is based on the RFC 6330 standard as explained in [19]. Basically, the RQ encoding process consists of three steps: 1) the creation of an $L$ x $L$ encoding matrix **A** (for a specific RQ block size $K$), 2) the generation of $L$ intermediate symbols by multiplying the original data with the encoding matrix, and 3) the generation of the encoding symbols by combining a small number of intermediate symbols. The resulting RQ encoded UDP packets then transmitted over the underlying Wi-Fi network.

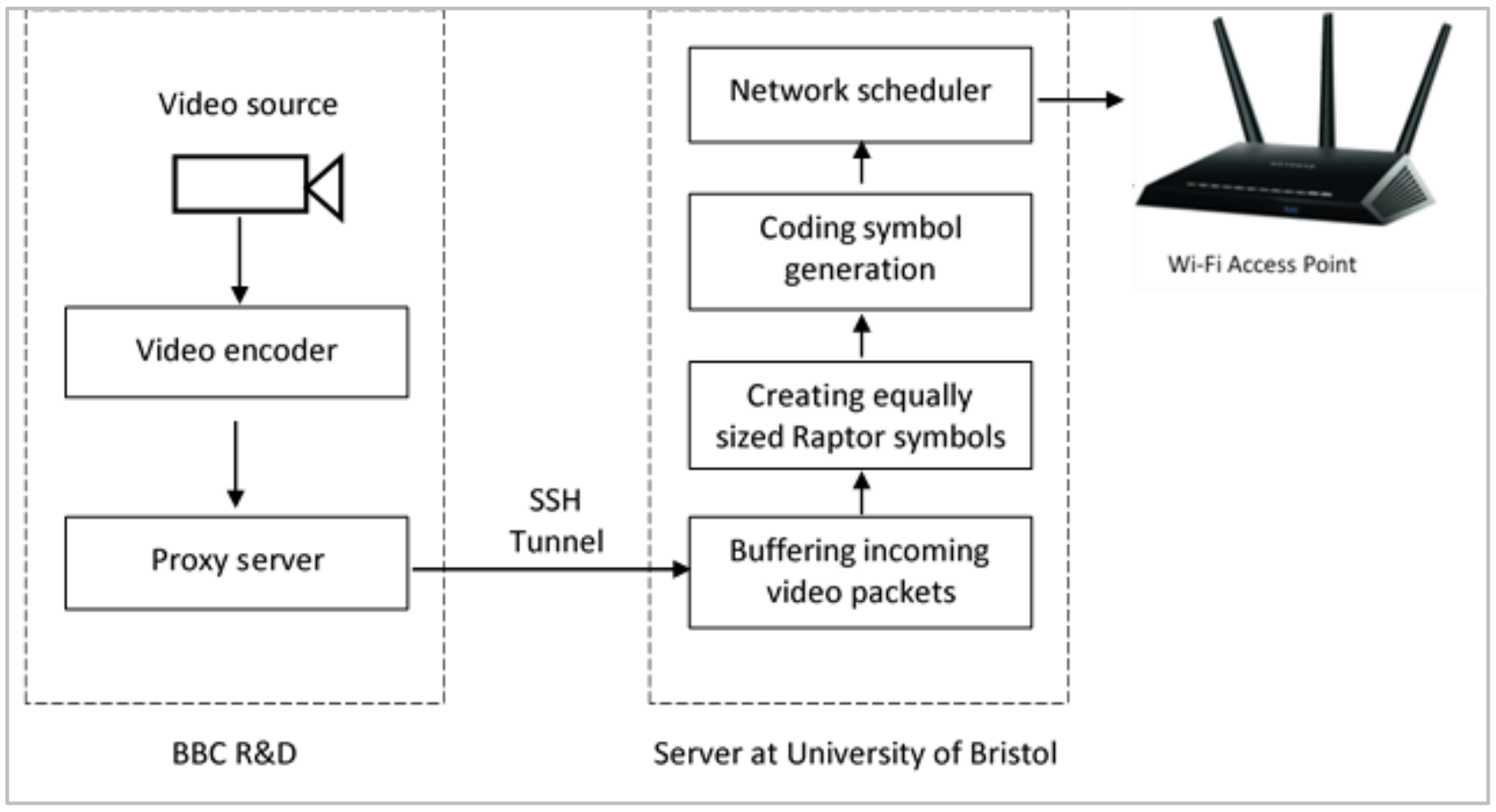

Fig. 2. Video streaming process at the transmitter.

## B. Client Configuration

The clients were Toshiba AT10LE Android tablets which have a single embedded antenna and connected to the AP at 2.4 GHz. Fig. 3 shows the process at the client. It is seen that at the client the received multicast packets are stored. If the number of received packets is $R \geq K'$, where $K'$ is the number of source plus padding symbols in an extended source block [19], then the RQ decoding is implemented over these symbols to recover the source block of the file. The first part of the decoding is the creation of an $L$ x $L$ encoding matrix A and then the generation of $L$ intermediate symbols by multiplying the received data with the encoding matrix. Finally, the $K$ source symbols are regenerated by employing LT (Luby Transform) decoding on intermediate symbols. If decoding is successful, then all $K$ packets are transmitted to video decoder otherwise only the received original packets are transmitted.

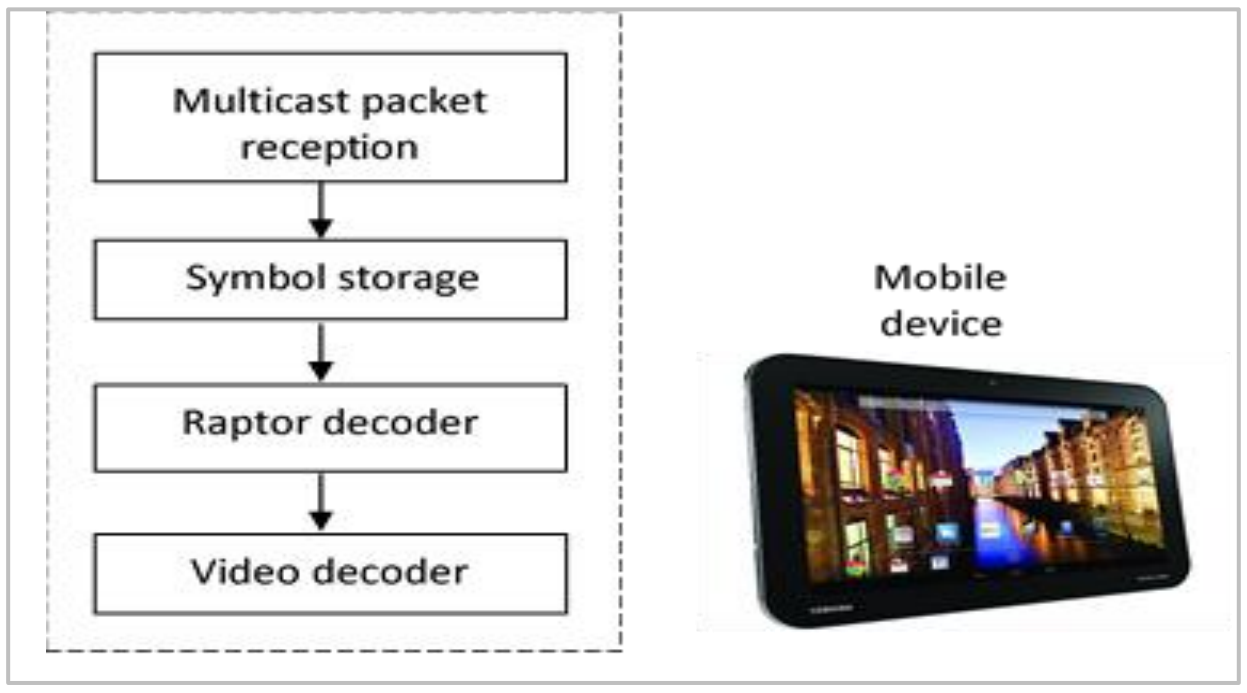

Fig. 3. Data processing at the receiver.

## C. *Measurement Scenario*

The measurements were conducted at the University of Bristol using two Toshiba AT10LE Android tablets and a single Netgear 7000 AP. The Netgear 7000 was flashed with internal Broadcom firmware to enable the selection of the MCS mode. The AP and the server were located at 12 m above the ground (on the 4th floor of the Merchant Venturers` Building) as shown in Fig. 4 and the server was connected to the AP via an Ethernet cable. The measurements were conducted at four different UE locations (distances) from the AP in Woodland Road, Bristol, UK. During the measurement runs, the users stood on average of 2 minutes at each of the UE location defined in Fig. 4 and logged the received packet traces and RQ source block size *K* values for each experiment (parameter set).

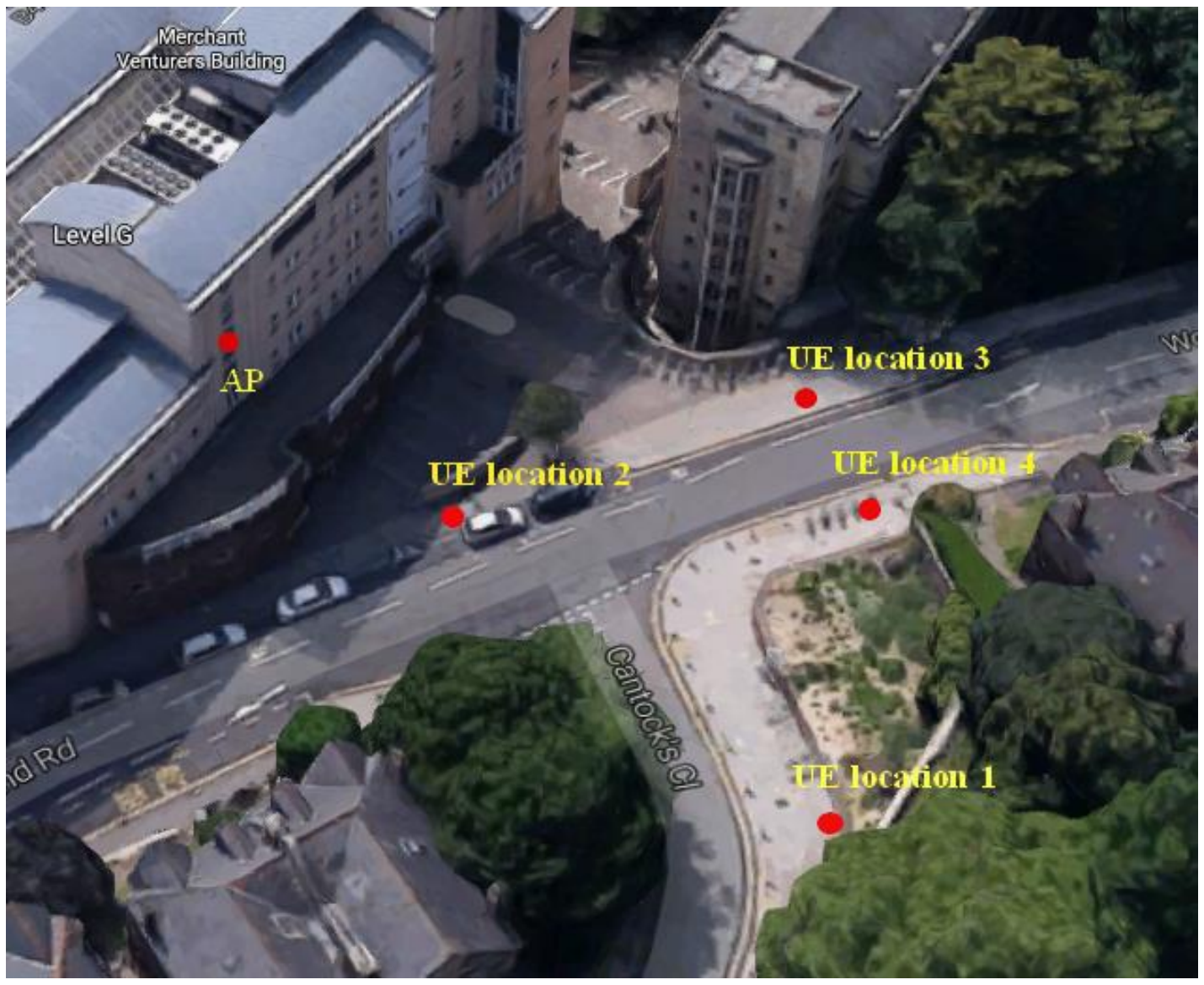

Fig. 4. Measurement scenario.

## D. *Measurement Parameters*

Detailed information on the measurement parameters is given in this section. An HD video steam provided by BBC was used during the measurement. In order to evaluate the UE performance, the incoming video stream was encoded using different RQ parameters such as source block size *K*, symbol size *T*, RQ code rate *CR* and buffering time $t_b$ as shown in Table I. In the PHY layer, the transmission modes for the IEEE 802.11n 20 MHz channel profile with an 800 ns guard interval as seen in Table II were used [1].

TABLE I. APPLICATION LAYER PARAMETERS

| Parameters | Value |
|---|---|
| Source block size, *K* | Variable |
| Source block window, *w* | 1 s |
| RQ symbol size, *T* | 500 B, 1400 B |
| Buffering time, $t_b$ | 5 s, 10 s |
| RQ code rate *CR* | 0.2, 0.33, 0.66 |

TABLE II. IEEE 802.11N TRANSMISSION MODES USED IN THE MEASUREMENTS

| MCS index | Spatial streams | MCS modes | Code rate | Data rate (Mbps), 20 MHz channel, 800ns GI |
|---|---|---|---|---|
| 0 | 1 | BPSK | 1/2 | 6.5 |
| 1 | 1 | QPSK | 1/2 | 13 |
| 2 | 1 | QPSK | 3/4 | 19.5 |
| 3 | 1 | 16-QAM | 1/2 | 26 |
| 4 | 1 | 16-QAM | 3/4 | 39 |
| 5 | 1 | 64-QAM | 2/3 | 52 |
| 6 | 1 | 64-QAM | 3/4 | 58.5 |
| 7 | 1 | 64-QAM | 5/6 | 65 |

## V. MEASUREMENT RESULTS AND ANALYSIS

This section presents the post-processed measurement results. The main performance evaluation metrics used to reflect the overall system performance (service quality) for live HD video streaming in multicast networks are the RQ decoding success rate and UDP PER before/after RQ decoding. Using the logged received packet traces and the source block size *K* values, the decoding success rate and the UDP PER were calculated for each set of parameters. The objective is to reduce the UDP PER in each source block and hence increase the RQ decoding success rate (user QoE).

### *A. Different MCS Modes*

The next set of results compare the RQ decoding success rate and UDP PER for different MCS modes and locations. In this experiment, the results are given for *T* = 1500 B, *CR* = 0.2 and $t_b$ = 10 s. It can be seen in Fig. 5 that increasing MCS modes decreases the decoding success rates. The reason is PERs in these MCS modes are very high thus the number of transmitted RQ repair symbols is not enough to compensate those packet errors. At location 3, the decoding success rates depending on the MCS show quite stable results due to the perfect antenna alignment. The results for the location 4 varies significantly because this location was susceptible to the blockage by passing by vehicles during the measurements. Although, the location 1 is also on the other side of the road the passing by vehicle could not block the signal as this location is far away from the road side. Fig. 6 shows the mean UDP PER for each MCS mode before and after RQ decoding. It is seen that without RQ coding none of the MCS modes can provide reliable multicast video streaming, i.e., UDP PER is higher than 1% which is assumed to be the maximum UDP PER value that can be tolerated by the video codec [20]. It is shown from the figures that the use of RQ codes enables lower MCS modes (MCS 0-3) to provide reliable video multicasting over Wi-Fi, i.e., UDP PER is less than 1%. It is also observed in Fig. 6 that even though a very low RQ code rate (*CR* = 0.2) was used, the RQ decoder was not able to successfully decode all the source blocks for some lower MCS modes with less than 0.5 mean UDP PERs before RQ decoding, i.e., residual UDP packet errors are observed at the application layer. In order to provide detailed insight, the UDP PER in each source block with/without RQ coding for MCS 3 at location 1 is presented in Fig. 7. It is seen that some source block experience severe error bursts as a result RQ decoder failed to decode the source blocks thus only received source packets were transmitted to higher layers. It should be noted that mean UDP PER results do not directly reflect the tablet performance (i.e., the end user QoE) since the RQ decoding success depends on the PER in each source block, rather than the average UDP PER.

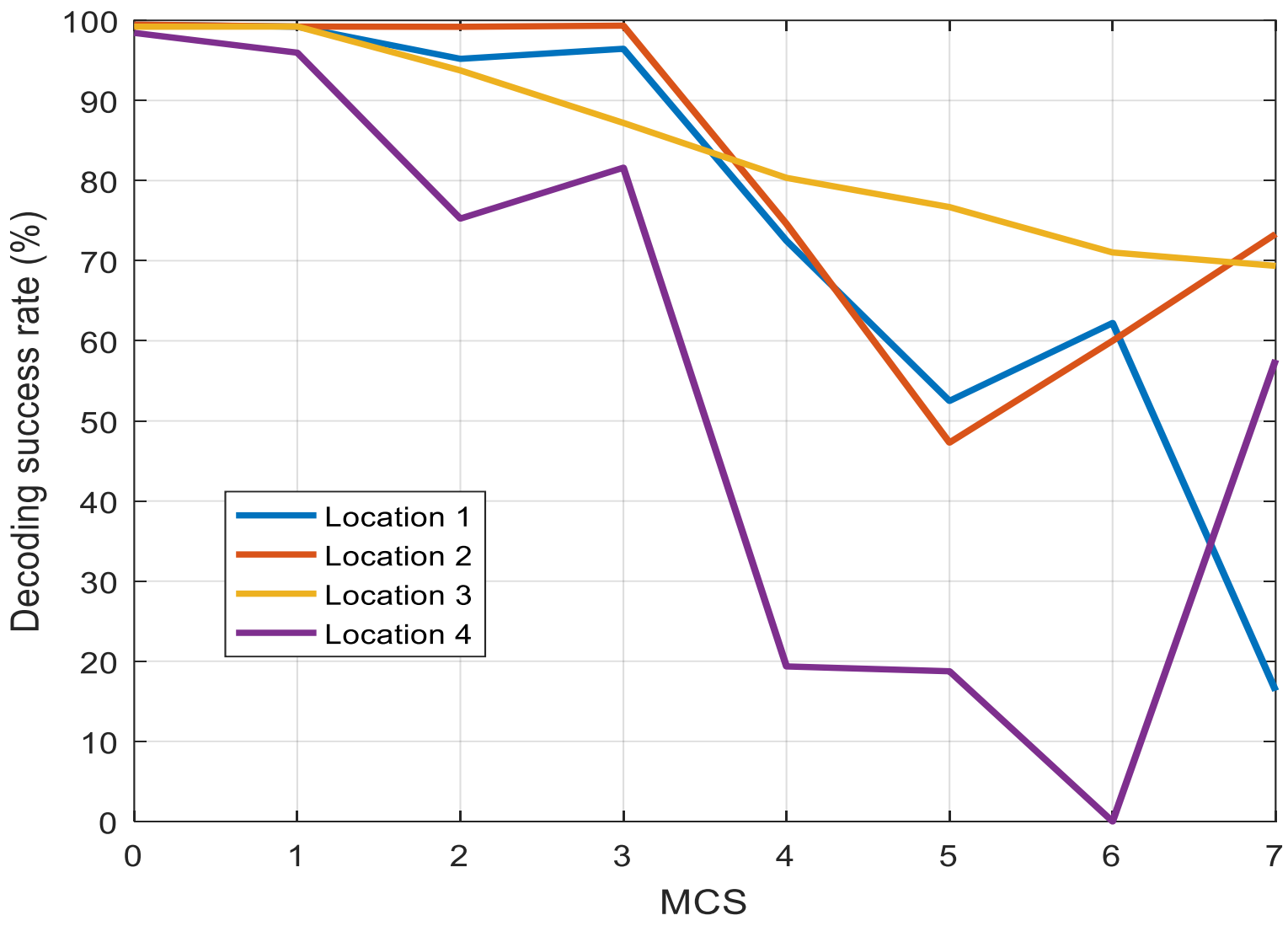

Fig. 5. Decoding success rate depending on MCS modes at each location.

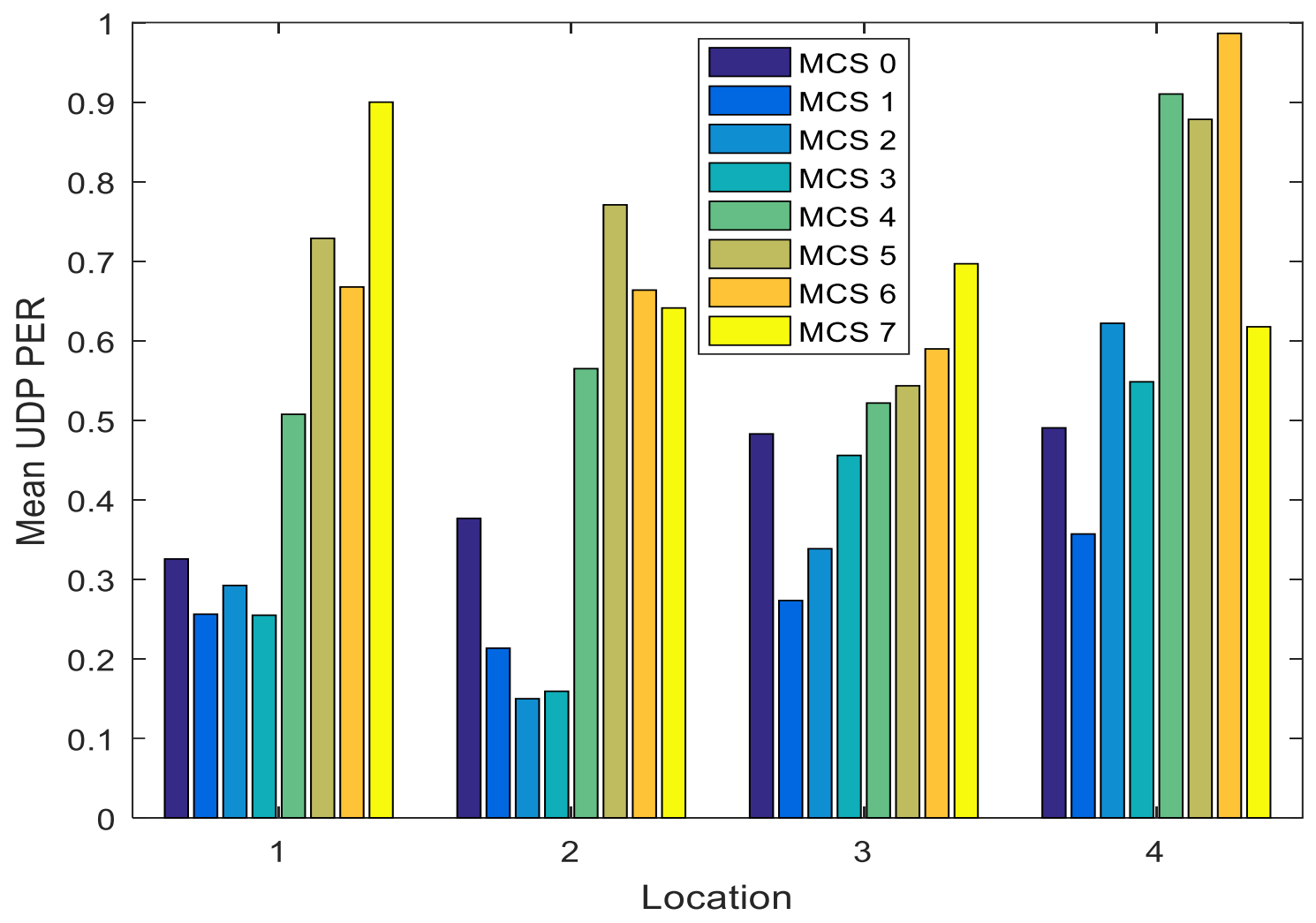

(a) Before RQ decoding.

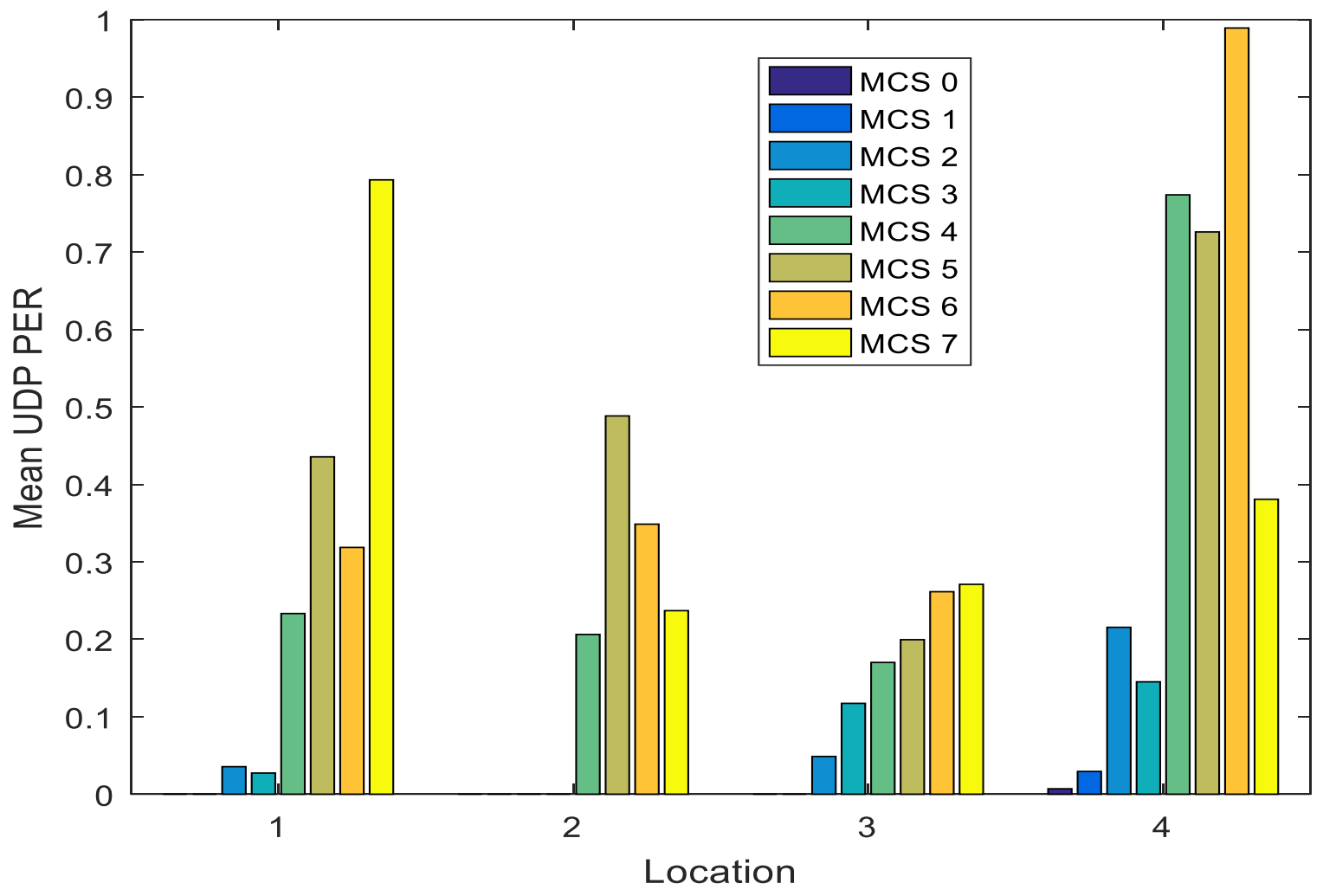

(b) After RQ decoding.

Fig. 6. Mean UDP PER depending on MCS modes before and after RQ decoding.

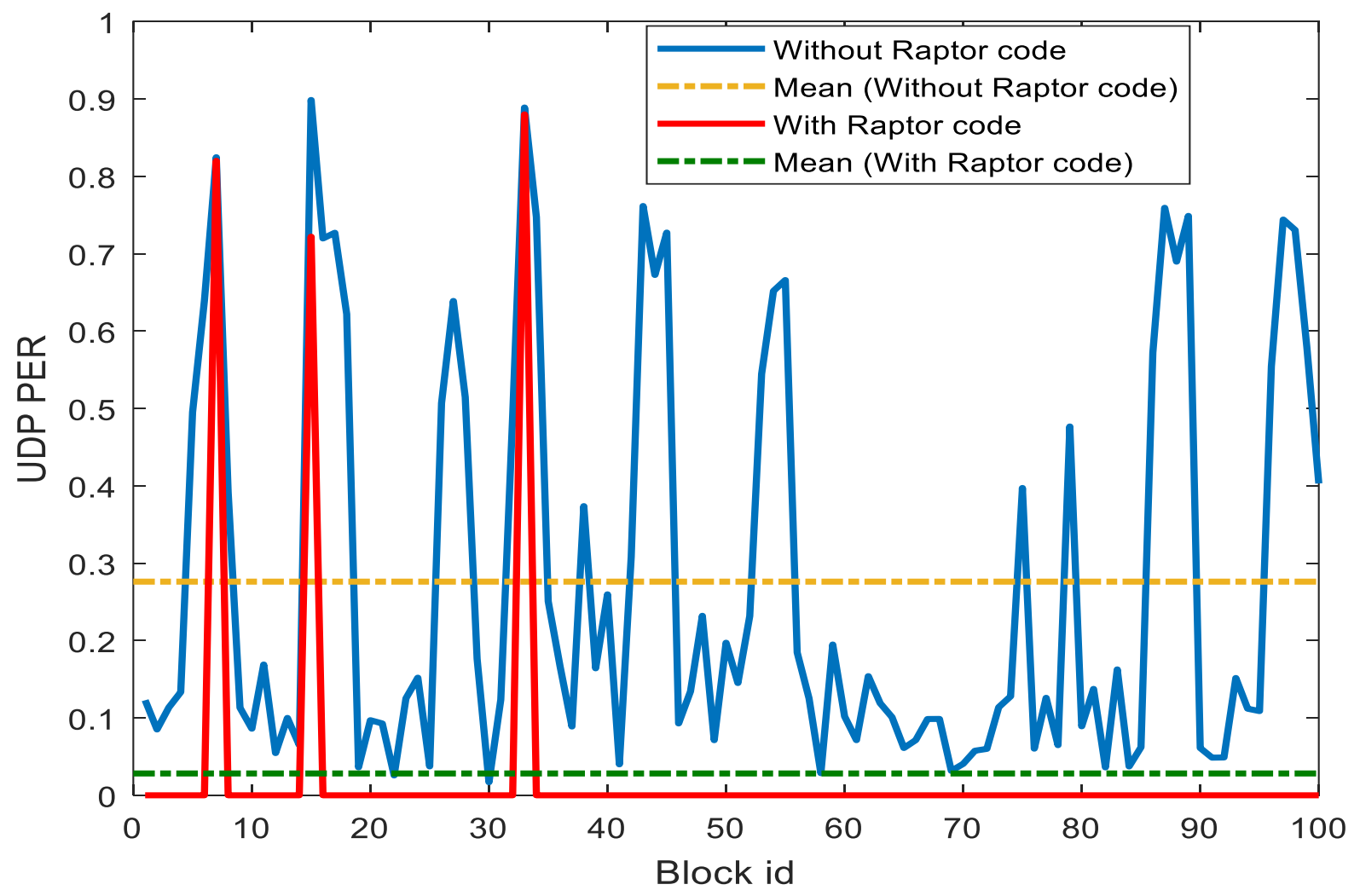

Fig. 7. UDP PER in each source block with and without RQ decoding for MSC 3 at location 1.

### B. *Different Buffering Times*

In this experiment, the decoding success rate is evaluated with respect to different buffering times $t_b$ of 5 s and 10 s for a fixed RQ source symbol size *T* of 1400 B and *CR* of 0.2. The buffering time is defined as the complete processing time of data including the coding, delivery and decoding times. Fig. 8 shows the results for MCS modes 1 and 3. It is observed that depending on the buffering time there is a slight change in the system performance, i.e., higher buffering time provides slightly better results as seen in location 1 and 2 and the variation in location 3 and 4 attributed the changes in the propagation environment (bad channel conditions as seen in Fig. 9).

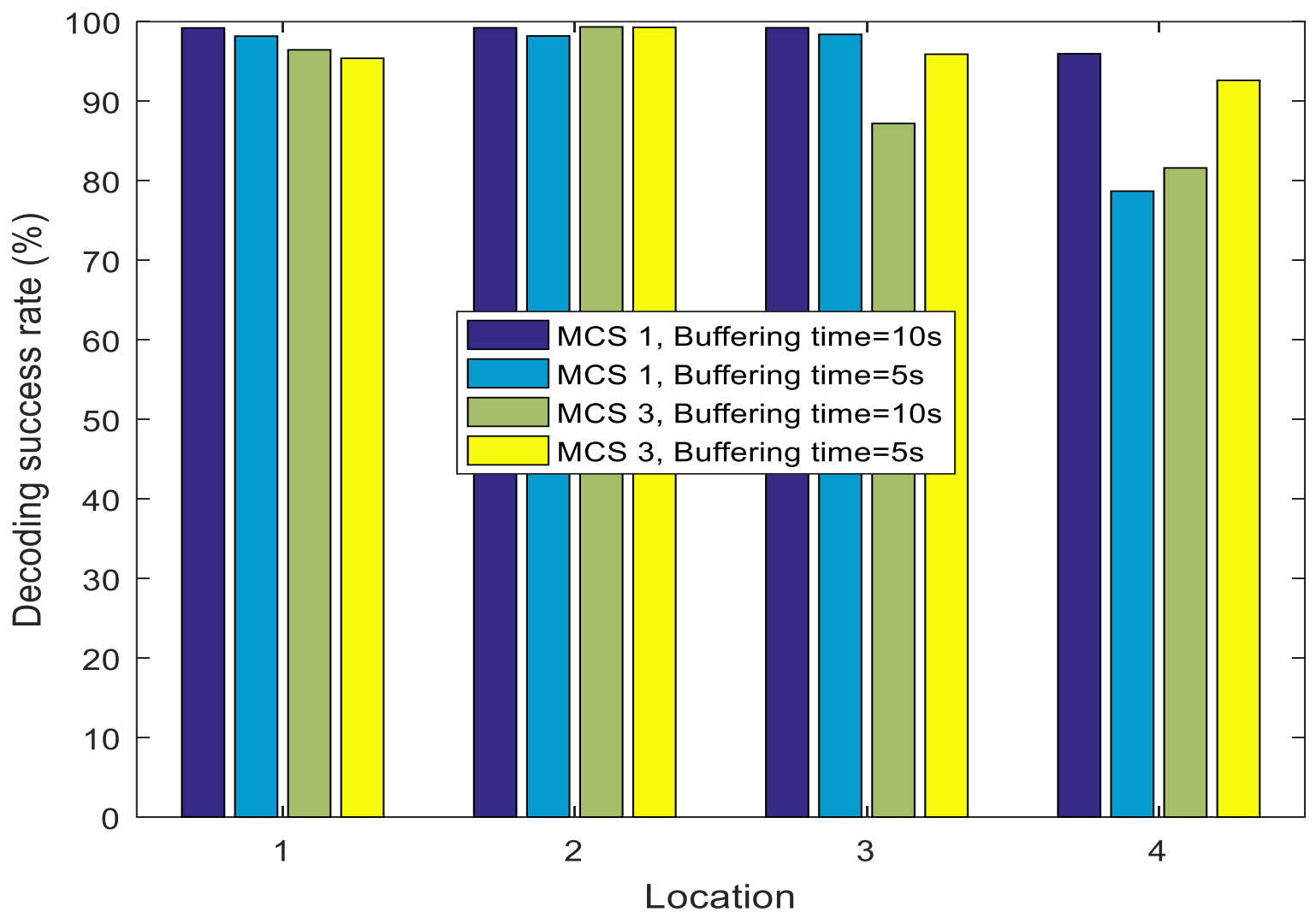

Fig. 8. Decoding success rate depending on buffering time for MCS 1 and MCS 3.

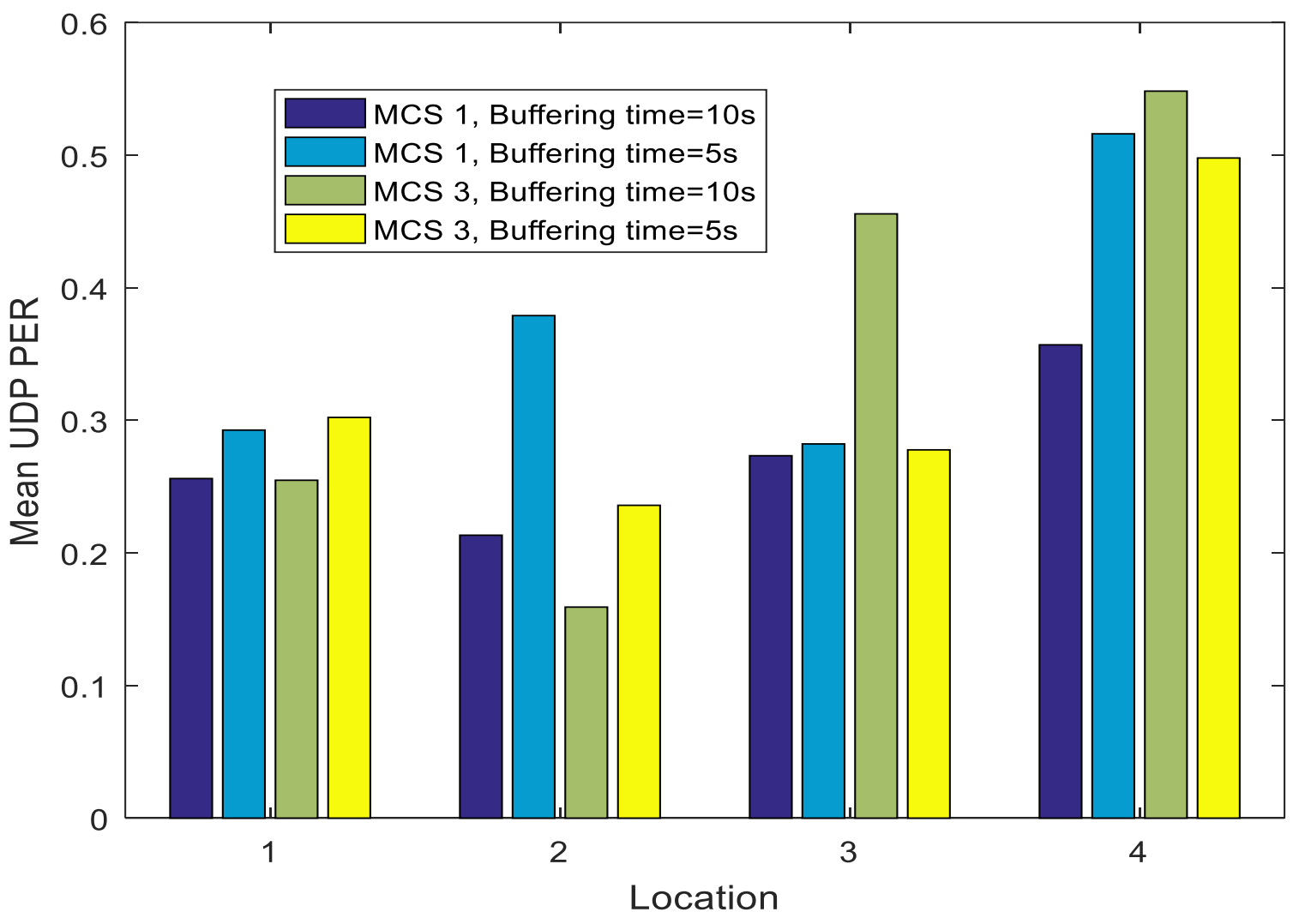

(a) Before RQ decoding.

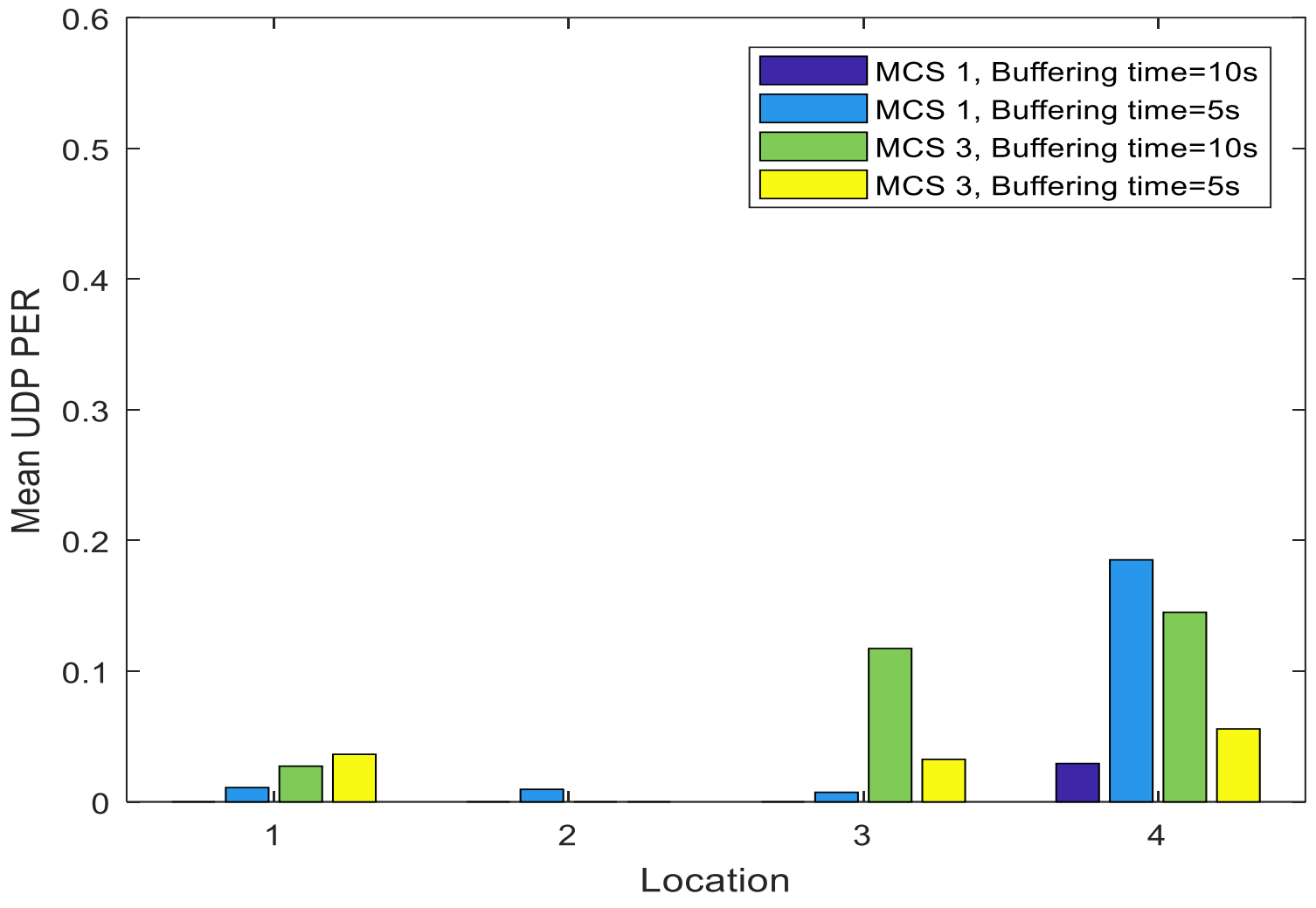

(b) After RQ decoding.

Fig. 9. Mean UDP PER depending on buffering time before and after RQ decoding.

### C. *Different RQ Symbol Sizes*

In this experiment, the decoding success rate is evaluated with respect to different RQ symbol sizes $T$ of 500 B and 1400 B for a fixed buffering time of $t_b$ = 10 s and a RQ code rate $CR$ of 0.2. Fig. 10 shows the decoding success rate for MSC 1 and 3. It can be seen that increasing the symbol size results in higher decoding success rates for locations with good channel conditions (locations 1-2). However, the use of small symbol size provides better user QoE for locations with bad channel conditions (locations 3-4), especially for higher MCS modes, as small PHY layer packets are less likely to be corrupted (i.e., PER is lower as seen in Fig. 11) compared to longer packets [15], [21]. Fig. 12 shows the UDP PER before RQ decoding for MCS 1 at location 1 and MCS 3 at location 3. It is seen that using large packet size results in longer burst of errors, especially for the higher MCS mode. It should be noted that reducing the symbol size leads to increase in the source block size $K$ since $K = \frac{wb}{T}$, where $b$ is the video bit rate, e.g., $K$ value ranges between 40-100 for $T$=1400 B however for $T$=500 B it ranges from 100 to 240. An increase in the $K$ value results in increased power consumption and delay at the transmitter and the receiver. Although this is not a big problem at the transmitter side it puts constraints on the mobile devices which have limited power and processing capabilities. Therefore, the symbol size should be selected considering the user QoE and the processing limitations of the mobile devices.

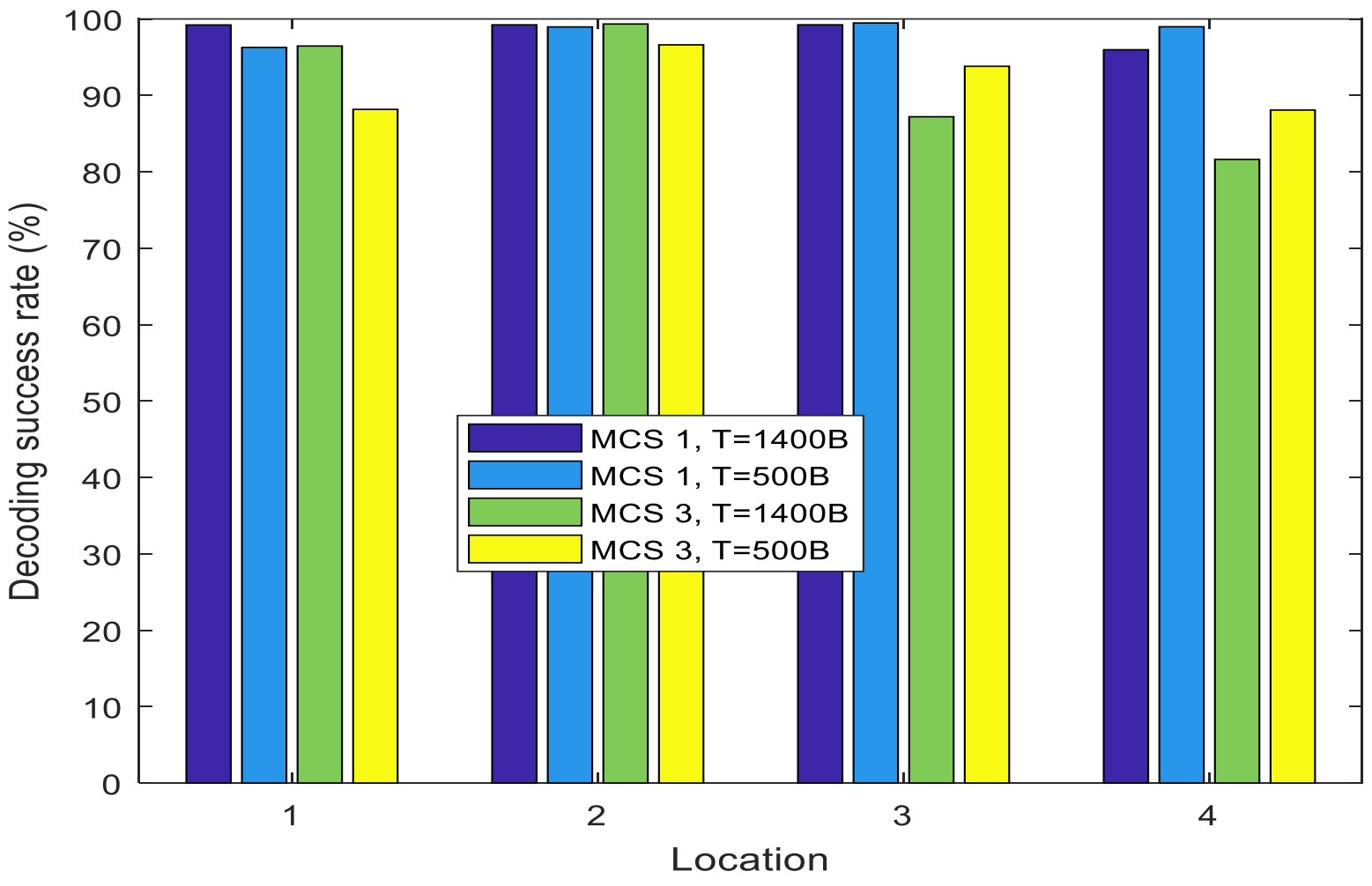

Fig. 10. Decoding success rate depending on RQ symbol size $T$ for MCS 1 and MCS 3.

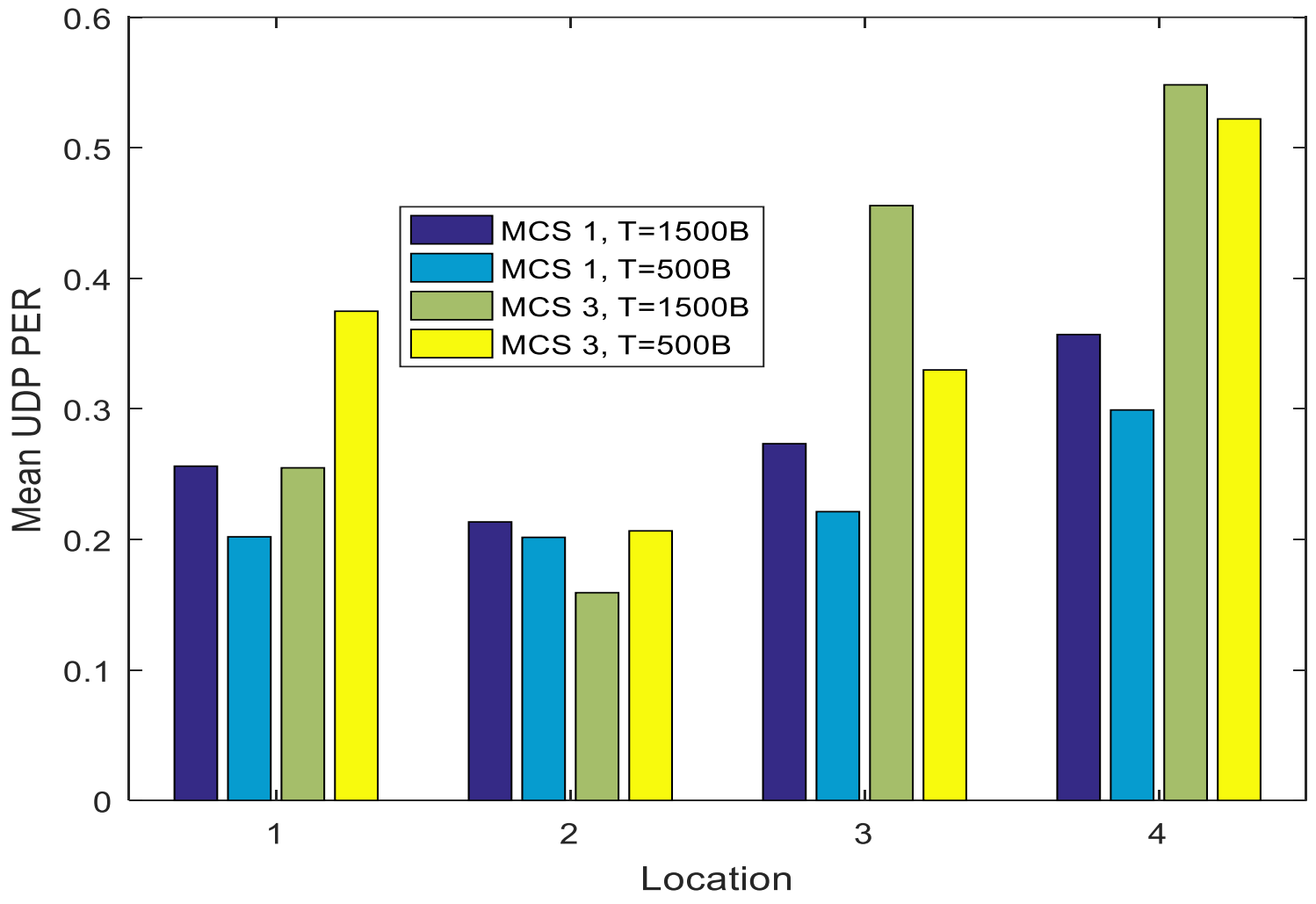

(a) Before RQ decoding.

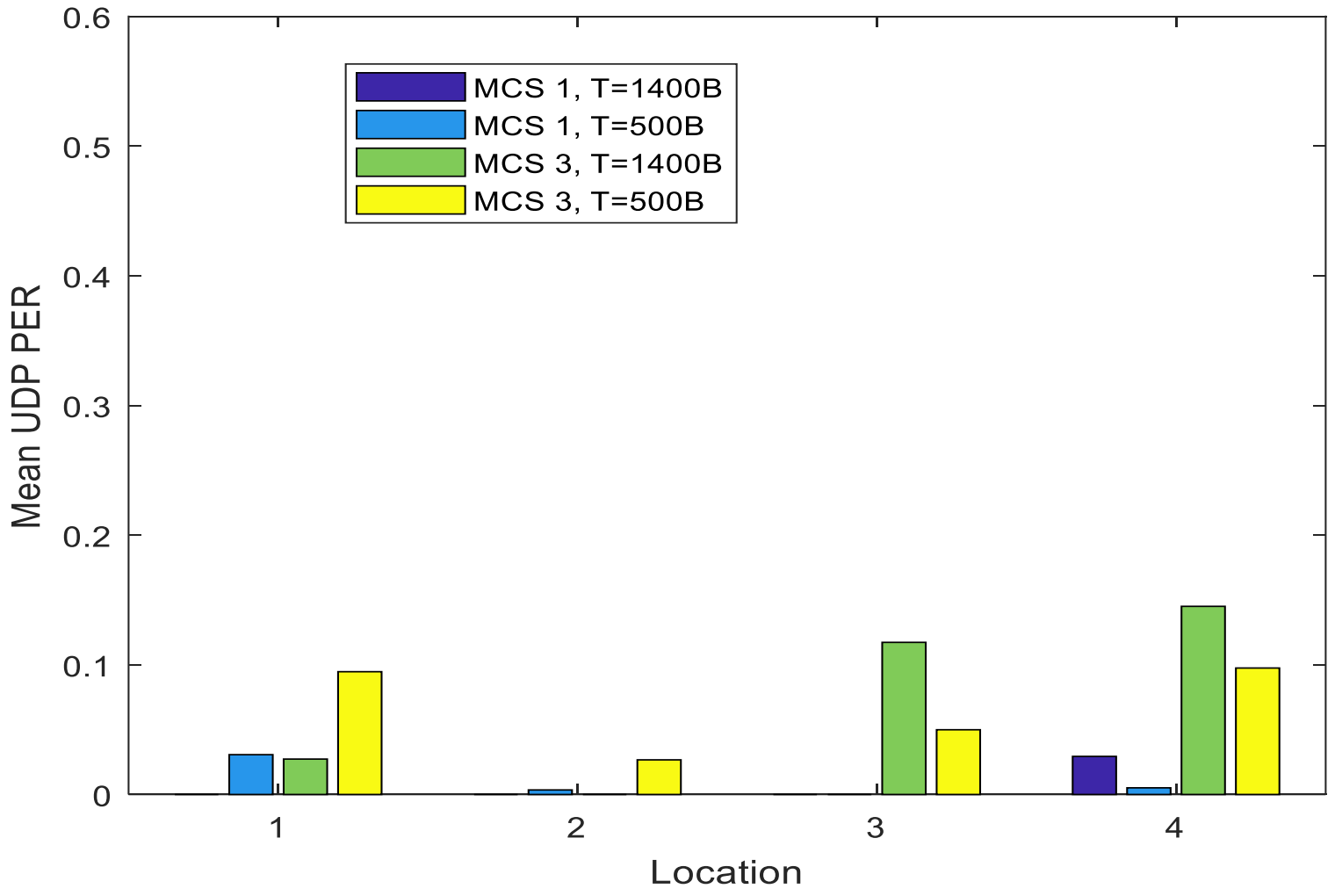

(b) After RQ decoding.

Fig. 11. Mean UDP PER depending on packet size $T$ before and after RQ decoding.

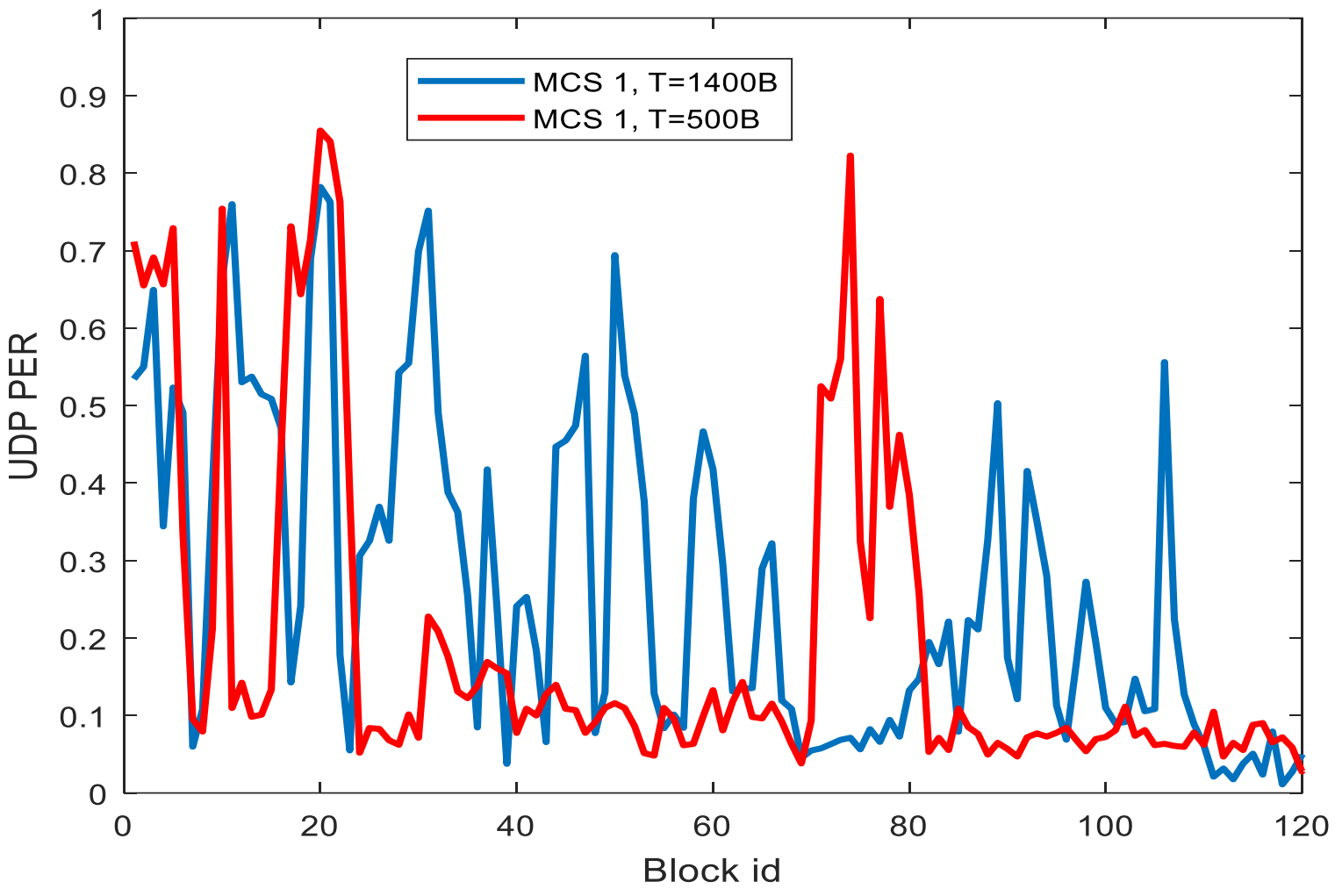

(a) MSC 1 at location 1

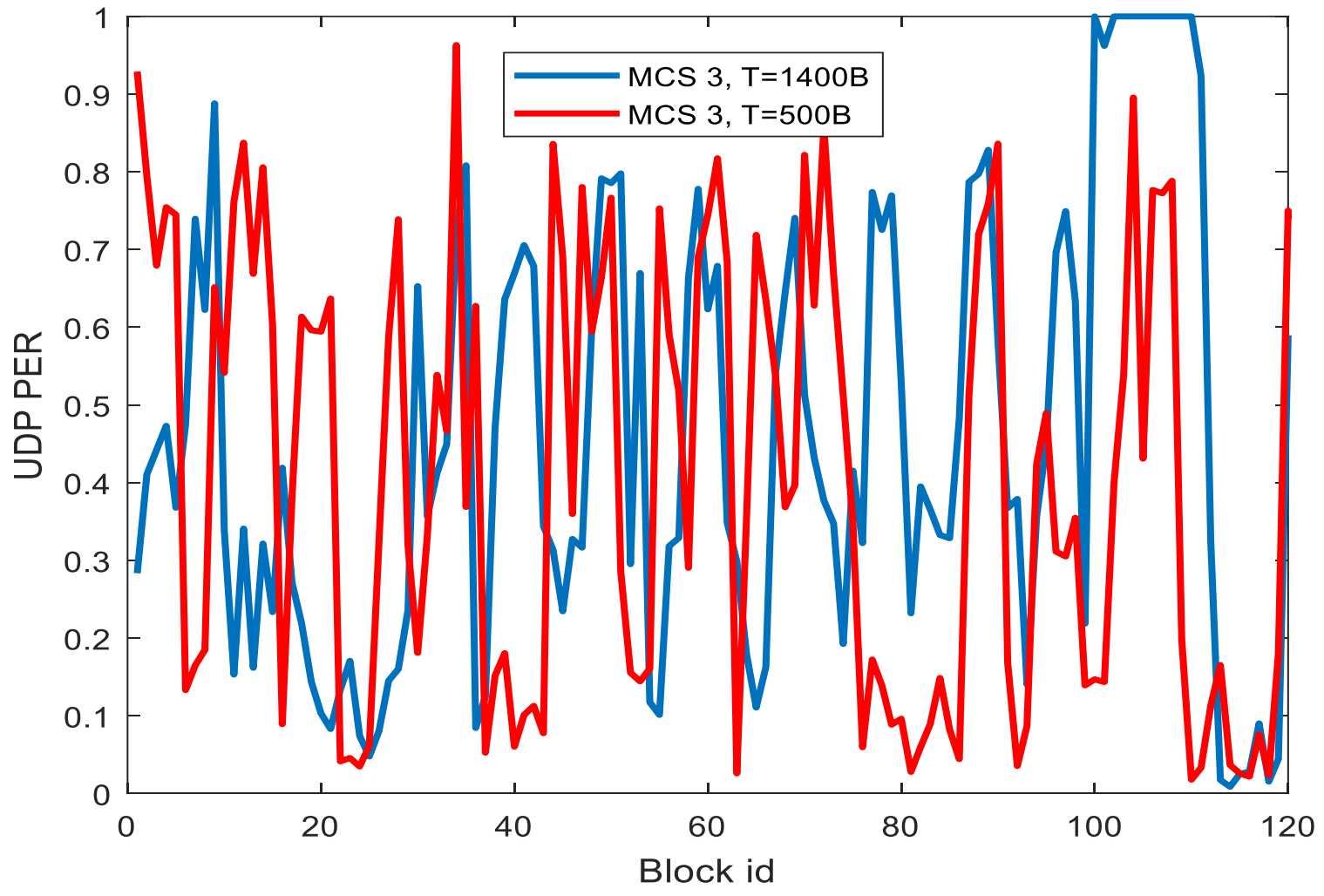

(b) MSC 3 at location 3

Fig. 12. UDP PER in each source block before RQ decoding.

### *D. Different RQ Code Rates*

In the last experiment, the decoding success rate is evaluated with respect to different RQ code rate *CR* of 0.2, 0.33 and 0.66 for a fixed buffering times of $t_b$=5 s and a RQ symbol size *T* of 1400 B. Fig. 13 shows the decoding success rate for MSC 1 and 3. It is seen that increasing the RQ code rates reduces the decoding success rates even for lower MCS modes since the number of repair symbols is reduced, i.e., there is not enough repair symbols to compensate burst of errors (seen in Fig. 14). As shown in Fig. 15 that PER in a source block can be as high as 88% therefore using a high *CR* of 0.66 cannot provide successful decoding. It is clear that for fixed code rate applications or deployments the code rate must be defined considering the experienced peak PER in source blocks. Although, this wastes valuable radio and network resources it provides reliable multicast video steaming, i.e., due to the bursty nature of the packet errors a lower code rate must be required for reliable multicast transmission over Wi-Fi. It is worth mentioning that when the *CR* reduces the packet arrival rate at the MAC layer increases due to the increase in the number of repair symbols. This leads an increase in the time-correlated packets errors as seen in Fig. 15 for example *CR*=0.2, i.e., it is more likely that the consecutive packets experience the same fading channel therefore the burst of errors increases.

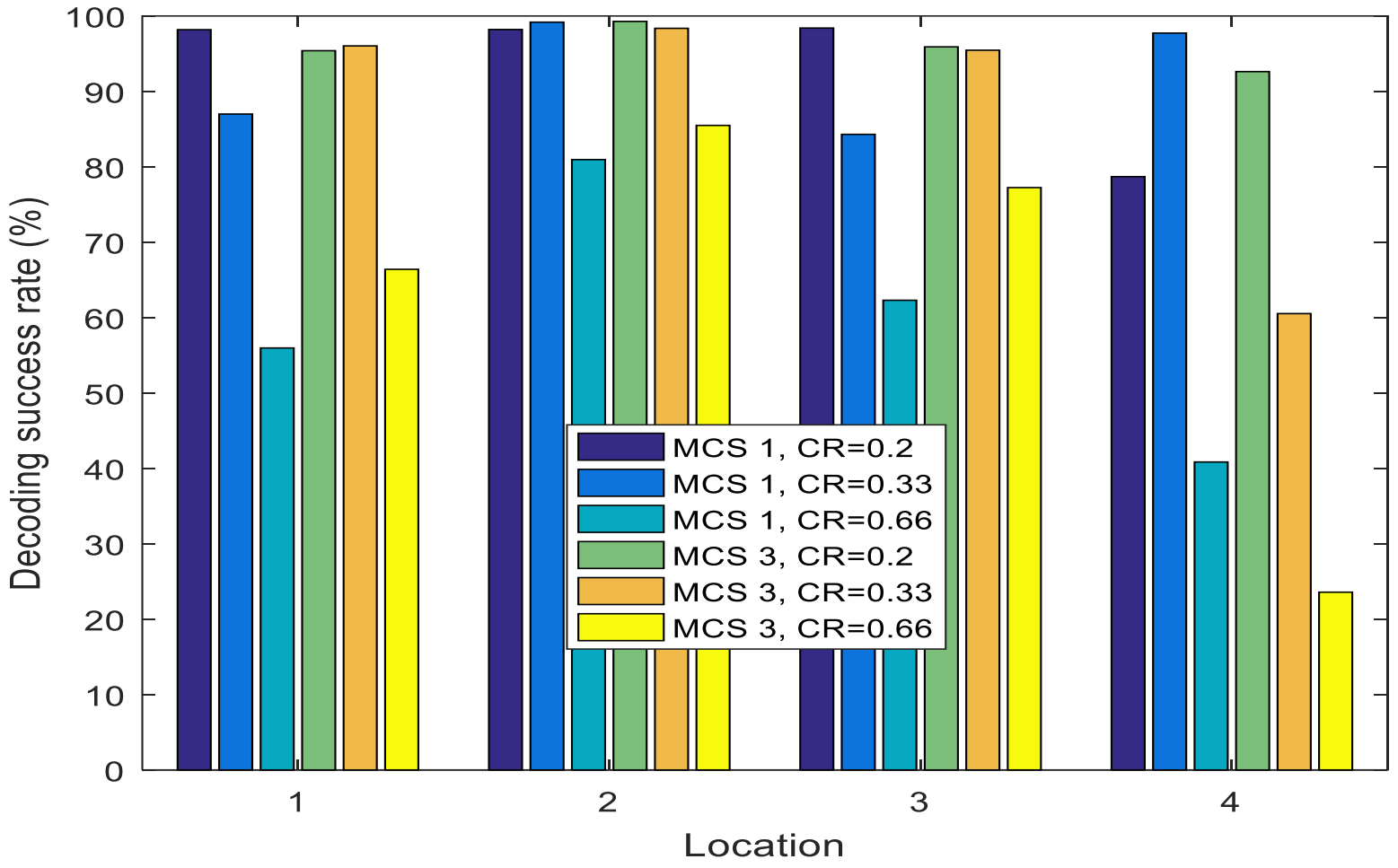

Fig. 13. Decoding success rate depending on RQ code rate *CR*.

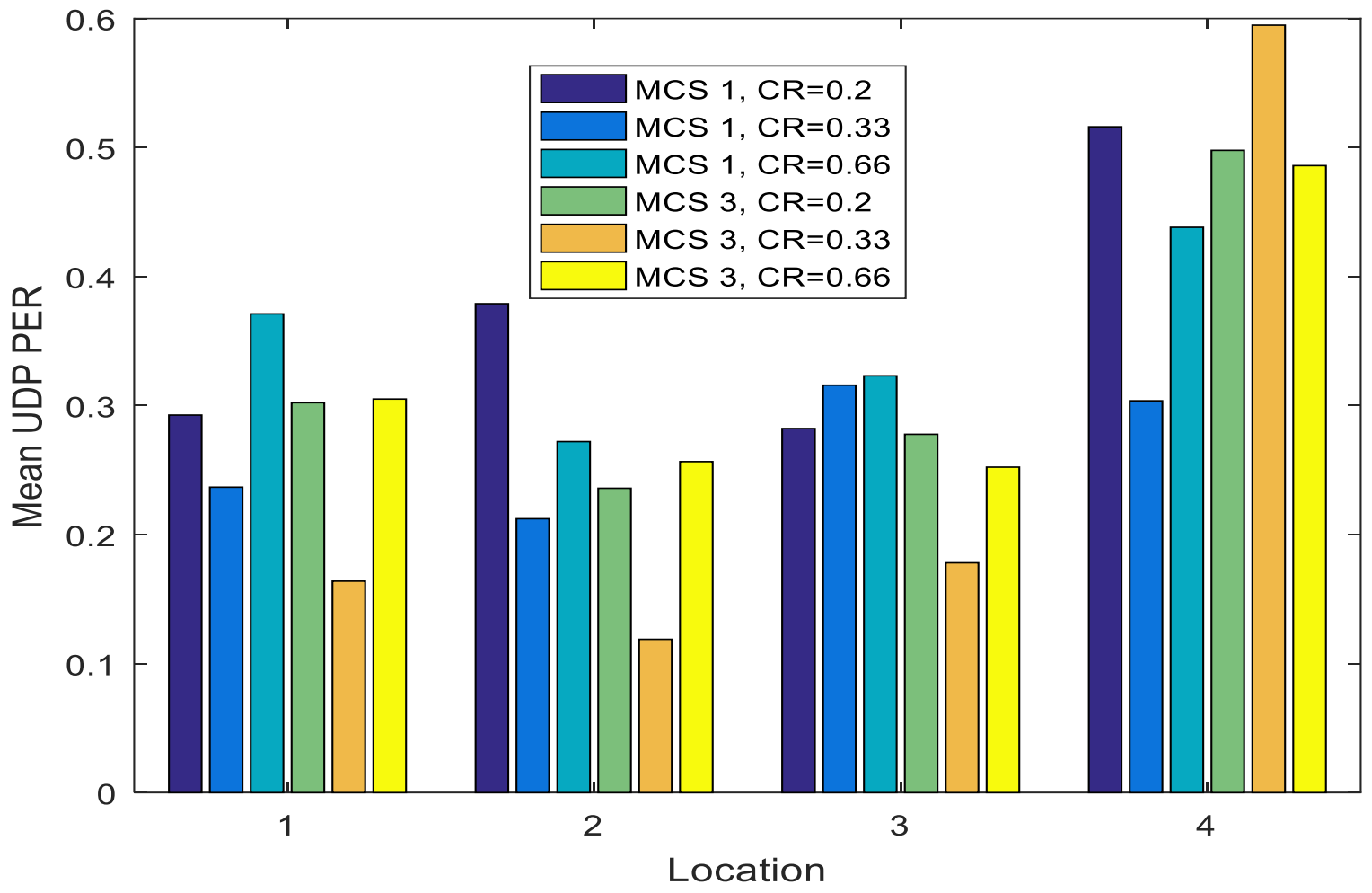

(a) Before RQ decoding.

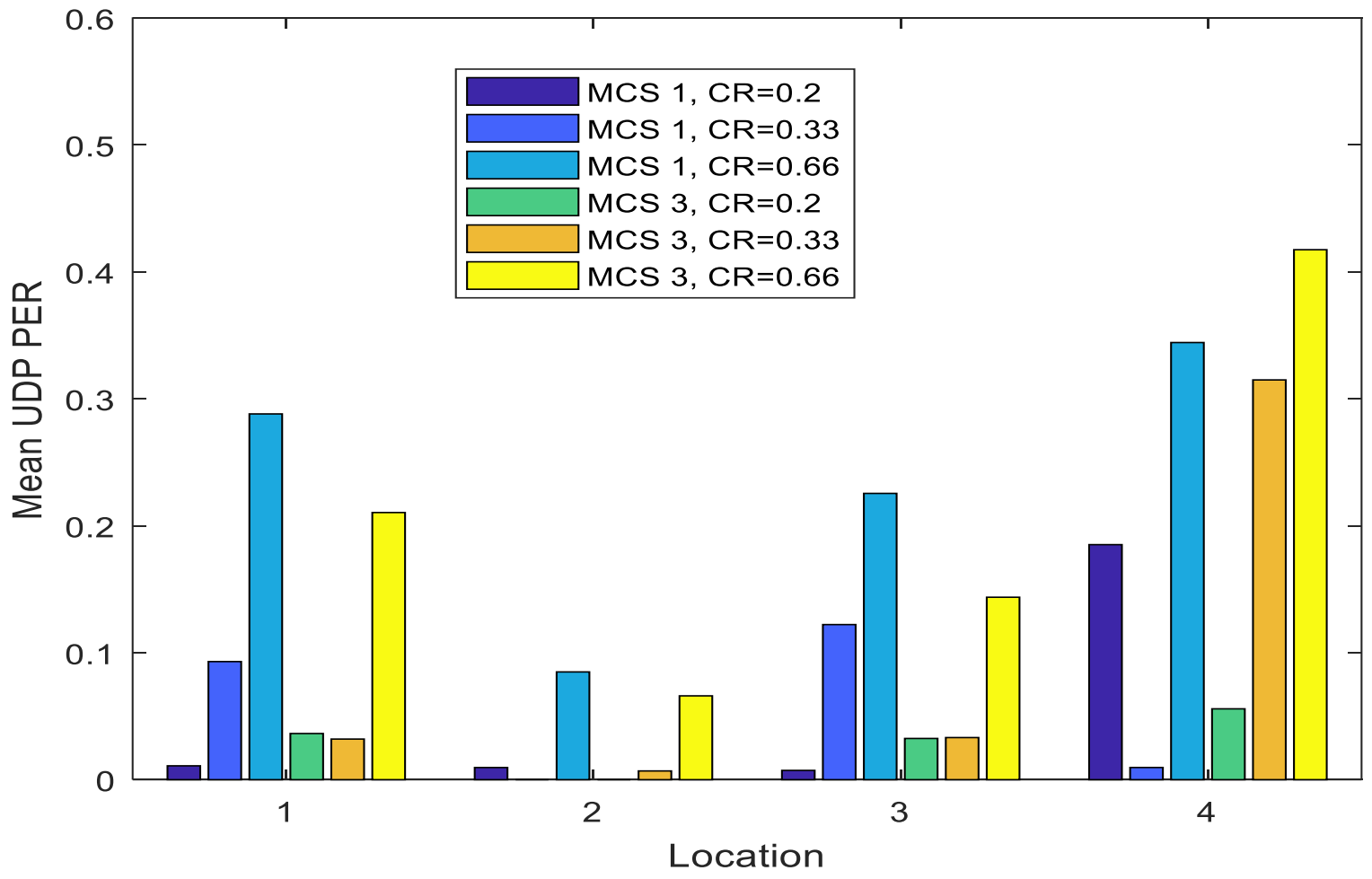

(b) After RQ decoding.

Fig. 14. Mean UDP PER depending on RQ code rate *CR* before and after RQ decoding.

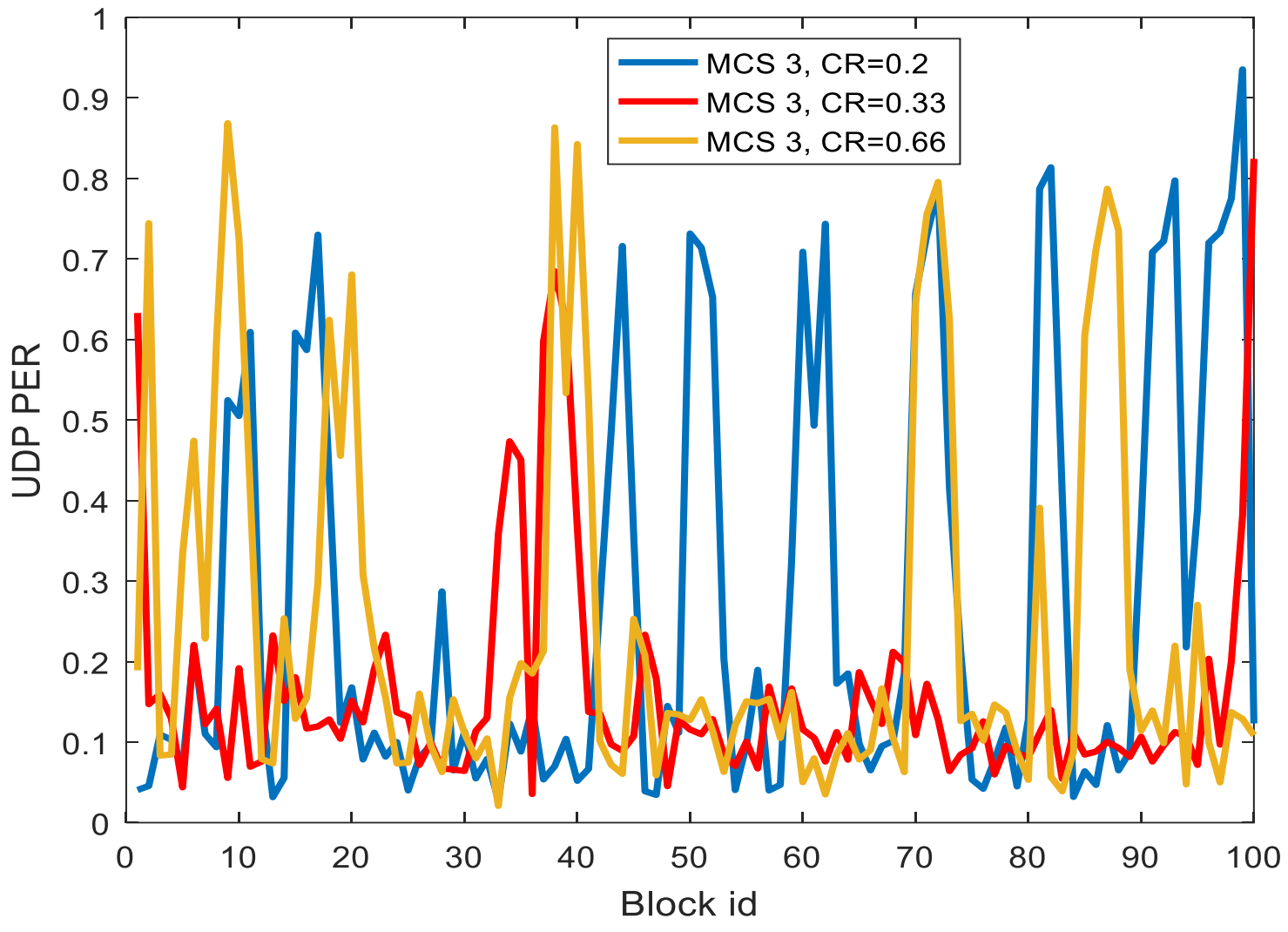

Fig. 15. UDP PER in each source block before RQ decoding for MSC 3 at location 3.

## VI. Conclusion

This paper presented a reliable and scalable wireless multicast HD video streaming solution for Wi-Fi considering AL-FEC RQ codes. A complete system design, implementation and evaluation process were presented. The results were reported in terms of real world measurements using a server, AP and tablet clients. RQ decoding success rate and UDP PER were used as key performance evaluation metrics to define user QoE and system performance. Detailed analysis on the implementation of RQ codes in a practical server/client system that transmit multicast HD video over Wi-Fi was presented. Results showed that system performance was mostly dominated by RQ code rate and the selected MSC mode. Due to the bursty nature of packet errors RQ code rate must be selected considering the maximum expected burst length in a source block. Furthermore, the results showed that the use of RQ codes enables lower MCS modes (MCS 0-3) to provide reliable video multicasting over Wi-Fi, i.e., UDP PER is less than 1%. Higher MCS modes require very low RQ code rate to provide reliable communications considering the cost of the radio and network resources that would not be very cost efficient. Moreover, video applications are very sensitive to the delay thus using a very low RQ code rate is not feasible in practice. When different RQ symbol sizes were compared, it was found that larger packet size provided better results than small packet, i.e., higher decoding success rates were observed for locations with good channel conditions. However, the use of small symbol size provided better user QoE for locations with bad channel conditions, especially when higher MCS modes used. As a conclusion, this paper provided a unique insight into the real-time implementation of RQ codes in a practical Wi-Fi multicast video streaming system. It was shown that scalable multicast video streaming services could be available over Wi-Fi using application layer RQ codes with suitable design parameters.

## ACKNOWLEDGEMENT

The author would like to thank Communication Systems & Networks (CSN) Research Group of University of Bristol for providing the measurement equipment and some CSN researchers for helping the measurements.